\documentclass[prb,article,twocolumn,superscriptaddress]{revtex4}

\usepackage{graphicx}
\usepackage{dcolumn}
\usepackage{bm}
\def\cm{cm$^{-1}$}

\begin{document}

\title{Charge-order fluctuations and superconductivity in two-dimensional
 organic metals}

\author{Alberto Girlando} \author{Matteo Masino}
\affiliation{Dipartimento di Chimica and INSTM-UdR Parma,
Universit\`a di Parma, Parco Area delle Scienze 17/a, I-43124 Parma, Italy}
\author{John A. Schlueter}
\affiliation{Material Science Division, Argonne National
Laboratory, Argonne, IL 60439-4831, and Division of Materials Research, National Science Foundation,
4201 Wilson Boulevard, Arlington, VA 22230, U.S.A.}
\author{Natalia Drichko}
\affiliation{
1.~Physikalisches Institut, Universit\"at Stuttgart, Pfaffenwaldring 57, D-70550 Stuttgart, Germany}
\affiliation
{Department of Physics and Astronomy,  Johns Hopkins University, Baltimore, Maryland 21218, U.S.A.}
\author{Stefan Kaiser}
\altaffiliation{Present address:
Condensed Matter Division,
Max Planck Research Dept. for Structural Dynamics at the University of Hamburg, CFEL,
Luruper Chaussee 149, D-22761 Hamburg, Germany}
\author{Martin Dressel}
\affiliation{
1.~Physikalisches Institut, Universit\"at Stuttgart, Pfaffenwaldring 57, D-70550 Stuttgart, Germany}


\begin{abstract}
We report comprehensive Raman and infrared investigations
of charge-order (CO) fluctuations
in the organic metal $\beta^{\prime\prime}$-(BEDT-TTF)$_2$SF$_5$CHFSO$_3$
and superconductor $\beta^{\prime\prime}$-(BEDT-TTF)$_2$SF$_5$CH$_2$CF$_2$SO$_3$.
The charge-sensitive vibrational bands are analyzed through
an extension of the well-known Kubo model for the spectral
signatures of an equilibrium between two states. At room temperature, both salts exhibit charge fluctuations between two differently charged molecular states with an exchange frequency of about $6\times10^{11}~{\rm s}^{-1}$.
The exchange rate of the metallic salt remains roughly constant down to 10~K, while
in the superconductor the exchange velocity starts to decrease below 200~K, and a ``frozen'' charge-ordered state emerges, and coexists
with the charge-order fluctuation state down to the superconducting temperature. These findings are confronted with other  spectroscopic
experiments, and a tentative phase diagram is proposed for the $\beta^{\prime\prime}$ BEDT-TTF quarter-filled salts.
\end{abstract}


\maketitle

\section{Introduction}

Understanding the mechanism of superconductivity
in strongly-correlated low-dimensional systems
represents one of the major challenges of modern solid state physics.
The competition/cooperation between different
interactions, including electron-phonon,
gives rise to complex phase diagrams, with Mott-like
instabilities close to the superconducting one.
In recent years, theoretical and experimental studies have
suggested that \textit{fluctuations} of an ordered state
may mediate superconductivity. Among organic superconductors,
magnetic fluctuations are thought to be relevant in certain
classes of compounds,\cite{powell11} but in recent years
attention has been focused on  charge-order (CO) fluctuations
occurring in quasi-two-dimensional quarter-filled
systems.\cite{merino01,dressel04,seo06}
However, the actual role of CO states and fluctuations
on the superconductivity mechanism
is still waiting for a deep understanding.

In order to shed light on the above issue, here we report
the temperature evolution of the infrared and Raman spectra
of two quarter-filled isostructural salts of
bis(ethylenedithio)tetrathiafulvalene (BEDT-TTF,
see inset of Fig. \ref{fig:beta_SC1}),
characterized by the so-called $\beta^{\prime\prime}$ packing.
One of them,  $\beta^{\prime\prime}$-(BEDT-TTF)$_2$SF$_5$CH$_2$CF$_2$SO$_3$
(hereafter $\beta^{\prime\prime}$-SC), is a superconductor with $T_c= 5$~K,
\cite{geiser96} whereas the other,
$\beta^{\prime\prime}$-(BEDT-TTF)$_2$SF$_5$CHFSO$_3$
(hereafter $\beta^{\prime\prime}$-M), remains metallic down to
the lowest temperature.\cite{ward00}

Infrared and Raman spectroscopy has indeed proved to be a very
useful tool in identifying the presence of charge disproportionation in
the organic compounds,\cite{bozio80,pecile89,dressel04} since
vibrational spectroscopy directly probes the charge distribution
on lattice sites. The frequencies of the charge-sensitive
vibrations of BEDT-TTF are known,  and probing of
charge distribution in the insulating state is a well established
technique.\cite{yamamoto05,girlando11}
Quite recently, it has been independently suggested\cite{girlando12,yakushi12}
that vibrational spectroscopy can also be fruitfully exploited
to analyze charges and charge fluctuations in metallic
systems. The idea is borrowed from nuclear-magnetic-resonance spectroscopy,
where it is well known that the line shape relevant to two
chemical species in equilibrium depends on the difference
in frequency between the two absorption energies and the speed
of exchange. In this ``two-states jump model'',\cite{kubo69}
when the exchange is slow with respect to the energy difference
one sees two bands, whereas when the exchange is fast one observes
a single band in between the two energy levels. In a previous paper,\cite{girlando12} we have
adapted the model to optical transitions, and tested it
on the charge-sensitive infrared active C=C antisymmetric stretch
of BEDT-TTF. Here we extend the model to Raman spectroscopy,
and present a full comparative analysis of the temperature
evolution of the spectra of the above mentioned
superconducting and metallic salts.

\section{Experimental}

The studied compounds were prepared as described in the
literature.\cite{geiser96,ward00} The infrared (IR) spectra
have been already published,\cite{drichko09,kaiser10} and are
reproduced here for comparison. The Raman spectra
have been  collected from the $ab$ crystal face with a Renishaw 1000 microspectrometer,
with 647.1 nm excitation from a Lexel Krypton laser, with 0.1 mW power at the sample.
Temperatures down to 10 K have been reached with ARS
closed cycle cryostat fitted under the Raman microscope.

\section{Vibrational selection rules and the effects of \lowercase{\textit{e-mv}} coupling}

\begin{figure}[h]
\centering
\includegraphics[width=0.88\linewidth]{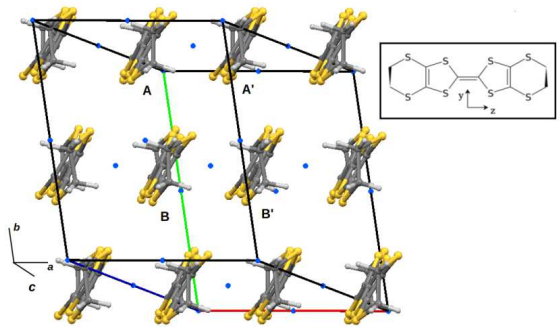}
\caption{Layer of BEDT-TTF molecules in $\beta^{\prime\prime}$-SC viewed
perpedicularly to the molecular long axis. The blue dots indicate
the inversion centers, A and B label the two
crystallographically independent molecular units. The organic
anion layers are not shown. Inset: structure of BEDT-TTF molecule,
with indication of the molecular axes.}
\label{fig:beta_SC1}
\end{figure}

The  $\beta^{\prime\prime}$-SC and  $\beta^{\prime\prime}$-M are isostructural,
crystallizing in the $P\overline{1}$ triclinic system, with two formula units per
unit cell.\cite{ward00, geiser96} The structure is characterized by layers
of BEDT-TTF in the $ab$ crystal plane, separated by the organic anions.
Within the cation layer, the BEDT-TTF are arranged in tilted dimerized stacks, typical
of the $\beta^{\prime\prime}$-packing motif,\cite{mori98} with the strongest
interaction along the crystallographic $b$-axis, i.e.\ perpendicular to the stacks.
The center of inversion
relates two BEDT-TTF molecules (e.g., A and A$^{\prime}$  in  Fig. \ref{fig:beta_SC1}),
and the two pairs, AA$^{\prime}$ and BB$^{\prime}$, are
crystallographically independent. We remark that the 300 K crystal structure of
$\beta^{\prime\prime}$-M exhibits some degree of disorder in the terminal ethylene groups.\cite{ward00}
 No disorder is observed in the structure of $\beta^{\prime\prime}$-SC, which however has been collected at 123 K.

For the Raman and IR selection rules  it is convenient to start by
dealing separately with the two pairs of inequivalent dimers
AA$^{\prime}$ and BB$^{\prime}$. Each pair is considered as a supermolecule,
residing on a crystallographic inversion center. We can therefore
apply the symmetric dimer model to carry out the
spectral predictions and to analyze the effects of the
electron-molecular vibration (\textit{e-mv})
interaction.\cite{rice79,painelli86} In a centro-symmetric dimer,
the vibrations of each molecular unit combine in-phase and out-of-phase.
In the $D_2$ molecular symmetry, the 72 normal modes of each BEDT-TTF
molecule are classified as: 18 $a$ (R) + 18 $b_1$ (R,IR) + 18 $b_2$ (R,IR)
+ 18 $b_3$ (R,IR), where R and IR indicate Raman and infrared activity, respectively.\cite{girlando11}
In the dimer, we then have 72 $A_g$ (R) and 72 $A_u$ (IR) vibrations.
The in-phase $A_g$ modes are completely decoupled from the charge-transfer (CT)
interaction within the dimer,\cite{rice79} and we expect weak
Raman intensity for modes originating from the in-phase combination of $b_1$,
$b_2$ and $b_3$ molecular vibrations. In other words, the Raman spectra of the
dimer will be dominated by the 18 totally symmetric ($a$) vibrations in the molecular
symmetry. The IR activity of the out-of-phase combination of the molecular
vibrations belonging to the $b_1$, $b_2$ and $b_3$ species will be substantially unaltered in the dimer:
the $b_1$ modes will be polarized along the $z$ long molecular axis, namely
approximately along the $c$ crystal axis (Fig.~\ref{fig:beta_SC1}),
the $b_3$ ones perpendicularly to the molecular planes (approximately along the $a$
crystal axis), and the $b_2$ perpendicularly to the former two (approximately along $b$).
On the other hand, the out of phase combinations of the $a$ modes interact with
the electronic system: they present a downward shift with respect to
the corresponding Raman active in-phase combination, and borrow intensity
from the CT electrons. Both the frequency shift and the intensity borrowing are
connected to  $\chi(0)$, the zero-frequency electronic susceptibility, and to
$g_i^2$, the square of the \textit{e-mv} coupling constants.\cite{rice79,painelli86}
The \textit{e-mv} coupled modes will be polarized along the stack ($a$ crystal axis),
like the $b_3$ molecular vibrations.

We now turn attention to the interaction between the AA$^{\prime}$ and BB$^{\prime}$ dimers.
It is formally an ``interstack'' interaction, but actually the hopping
integral between the stacks is larger than that along the stack.\cite{ward00}
We shall continue the analysis by taking the dimer as a supermolecular
entity, although the results are not different from those obtainable
by starting from the separated molecules. In the following, when we refer
to modes of $b_1$ symmetry, for instance, we actually mean ``the combination of molecular
$b_1$ modes within the dimer (either in-phase or out-of-phase, depending
on the context). Since the dimers are inequivalent, we can speak about
approximately in-phase and out-of-phase coupled dimer vibrations,
with corresponding pseudo Davydov splitting.\cite{turrell} In any case,
the spectral prediction concerning the non-totally symmetric (in the molecular
symmetry) vibrations are not substantially altered with respect to
those seen above for a single dimer (or stack). For example, the $b_1$ modes
we shall be concerned with in the following, are weak in Raman,
and appear in the IR spectra polarized along the $c$-axis (almost
perpendicular to $b$). They will not exhibit any appreciable splitting, since anti-parallel
transition dipole moments cancel each-other.\cite{schwoerer}
On the other hand, important spectroscopic effects are expected
for the $a$ modes, following the ``interstack'' CT interaction,
due to the \textit{e-mv} coupling. In the analysis, we can follow
again a dimer-like approach. In this case, however, the dimer
is non-symmetric: according the x-ray data, the A and B stacks
are inequivalent and bear a different molecular charge.\cite{geiser96,ward00}
The consequences of \textit{e-mv}
coupling on the IR spectra of such kind of systems has been first
analyzed by M. J. Rice \textit{et al.}\cite{rice80} for MEM(TCNQ)$_2$
(MEM = N-methyl-N-ehylymorpholinium, TCNQ = tetracyanoquinodimethane). 
Girlando \textit{et al.}\cite{girlando82} have extended the model to consider the consequences
of \textit{e-mv} coupling on the Raman spectra of a mixed-stack
solid like TTF-CA (TTF = tetrathiafulvalene,
CA = chloranil), which from the point of view of symmetry
is equivalent to a non-symmetric dimer.

The spectral consequences of \textit{e-mv}  coupling in
a non-symmetric dimer differ from those in a symmetric dimer.
In particular, there is no in-phase and out-of-phase
coupling between monomer-degenerate modes. This has an important consequence
for the mutual exclusion rule for the Raman and infrared activity: all the $a$ modes
of the BEDT-TTF moieties are active (with the same frequency) in Raman and in IR,
hence both the spectra are affected by the \textit{e-mv} interaction.
The frequencies of the \textit{e-mv} perturbed modes can be obtained from
the diagonalization of the following ``force constants'' matrix:\cite{painelli86,hanfland88}
\begin{equation}
F_{ij} = \omega_{i}\omega_j\delta_{ij} - \chi(0) \sqrt{\omega_i\omega_j} g_i g_j /\hbar
\label{e:forcematrix}
\end{equation}
where $\omega_{i,j}$ is the unperturbed frequency, and $\delta_{ij}$ is
the Kronecker delta. The zero-frequency electronic susceptibility depends
on the CT excitation frequency, $\omega_{\textsc{ct}}$, and on
the molecular ionicity $\rho$, $\chi(0) = 4\rho (1-\rho) /\omega_{\textsc{ct}}$,
and is maximum for equally charged molecules (the symmetric dimer).
We finally remark that whereas the consequences of \textit{e-mv} interaction
on the IR spectral features have been accurately reproduced in terms of the
appropriate model,\cite{rice80,girlando82} very little is known about the
expected Raman intensities, also because these are connected more to the
intra-molecular electronic structure.

\begin{figure}[ht]
\centering
\includegraphics[width=0.95\linewidth]{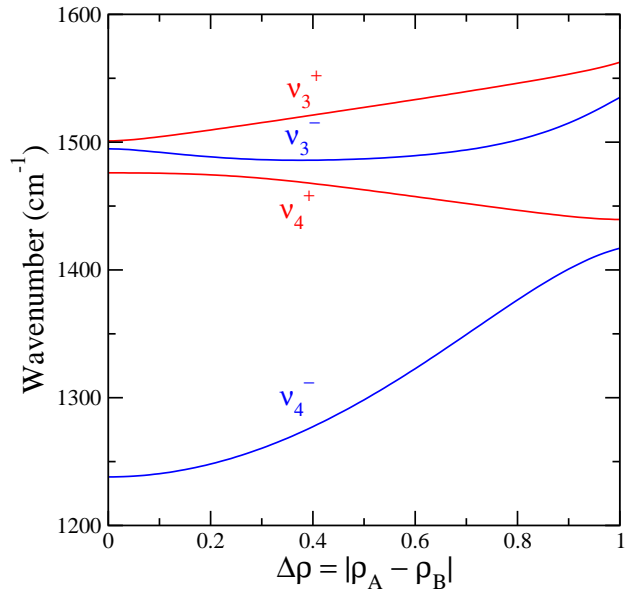}
\caption{Frequencies of the two charge-sensitive BEDT-TTF
C=C stretching $a$ modes ($\nu_3$ and $\nu_4$) as a
function of the charge difference $\Delta \rho$ between the two
inequivalent stacks. The red and blue curves correspond to the
charge-poor ($\rho < 0.5$) and charge-rich ($\rho > 0.5$)
molecules, respectively.}
\label{fig:ETrho_2}
\end{figure}

In the following, we shall mainly deal with the charge-sensitive
molecular vibrations, namely the three C=C stretching modes,
$a~\nu_3,~a~\nu_4$, and $b_1 \nu_{22}$
(known respectively as $a_g \nu_2,~ a_g \nu_3$, and $b_{1u} \nu_{27}$
in the $D_{2h}$ molecular symmetry).\cite{girlando11}
The $b_1 \nu_{22}$ mode is not coupled to the electronic system.
It may appear with weak intensity in the Raman spectrum, but
is well detectable in the IR spectra polarized along the $c$-axis.
In case the A and B dimers (on adjacent stacks) bear different charges, it will
appear as a doublet, with the same polarization.
The $a$ modes are instead affected by the \textit{e-mv}
interaction, connected to the
interaction between the inequivalent dimers. Therefore
the frequencies of the $a$ modes may have a nonlinear dependence on the molecular
charge. Such dependence can be calculated by diagonalizing
the $4 \times 4$ matrix of Eq.~(\ref{e:forcematrix}),
and is shown in Fig.~\ref{fig:ETrho_2}. The parameters used are:\cite{girlando11}
$\omega_3 = (1501 - 123~\rho$) \cm , $g_3 = 43$ meV, $\omega_4 = (1476 - 118~\rho$) \cm ,
$g_4 = 71$ meV. In the spirit of a semiempirical approach,
the CT frequency $\omega_{\textsc{ct}}$ which appears
in the espression $\chi(0)$ in the dimer model has been taken as adjustable
parameter, $\omega_{\textsc{ct}} = 2000$ \cm.

In Fig.~\ref{fig:ETrho_2} the frequencies of the $a~\nu_3$ and $\nu_4$ modes
of the two stacks are displayed as a function of their charge
difference, $\Delta \rho = |\rho_\mathrm{A} - \rho_\mathrm{B}|$.  $\Delta \rho = 0$
corresponds to two equivalent stacks, with uniform charge distribution, and in this point the difference between the
in-phase (red in the figure) and out-of-phase (blue) coupled vibrations
reflects the only effect of \textit{e-mv} interaction. By increasing $\Delta \rho$
above zero all the four vibrations become coupled together and to the electronic system
with the resulting non-linearities observed in the Figure.
The red and blue curves now correspond to charge-poor and charge-rich molecules,
respectively. Actually, as it has been already noted,\cite{yakushi12}
the $a \nu_3$ mode retains a linear behavior for small charge disproportionation
$\Delta \rho \lesssim 0.2$.
Above such a values, only the frequency of the ``charge poor'' molecule
($\rho < 0.5$, $\nu_3^+$ red line) continues to have a linear dependence. In any case, 
the molecular ionicity can be estimated  for the whole range of values by comparing the
observed Raman frequencies to Fig.\ref{fig:ETrho_2} diagram.

\section{Results}

\begin{figure}[ht]
\centering
\includegraphics[width=0.9\linewidth]{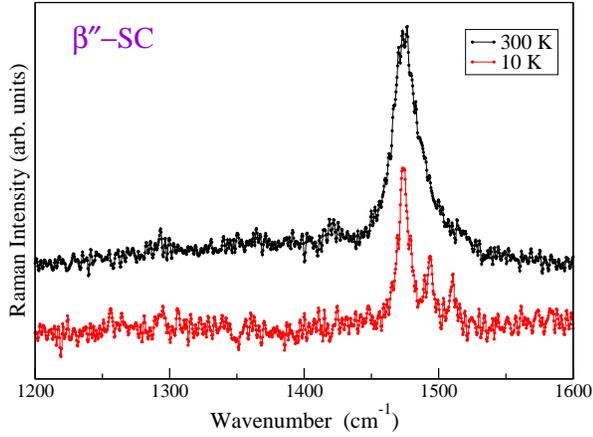}
\caption{Raman spectra of $\beta^{\prime\prime}$-SC from 1200 to 1600~\cm\ at $T=300$ and 10~K.
The spectra are offset for clarity.}
\label{fig:ramanSC-300_10}
\end{figure}

Fig.~\ref{fig:ramanSC-300_10} reports the Raman spectra of
$\beta^{\prime\prime}$-SC at room temperature and at 10 K in the spectral
region where the totally symmetric charge-sensitive
modes occur (1200-1600 \cm, cf. Fig.~\ref{fig:ETrho_2}). At 300 K,
only one asymmetric band is detected, which splits into three
bands as the temperature is reduced to 10~K. No band is detected in the region
where the $a~\nu_4^-$ mode
is expected. The $a~\nu_4$ is the mode most strongly coupled to the
electronic system, and the $a~\nu_4^-$ component displays large IR
intensity,\cite{girlando11} but is generally rather weak
and broad in Raman. Thus it is rarely detected there.\cite{yamamoto12} In the present
case the mode is difficult to identify also in IR, since it overlaps
with low-energy electronic transitions.\cite{kaiser10,dong99}
The three bands seen in the low-temperature spectrum
of $\beta^{\prime\prime}$-SC clearly correspond to the $a~\nu_3^+,~\nu_3^-$
and $\nu_4^+$ (in the order of decreasing wavenumber).
The latter is rather insensitive to the
charge (Fig.~\ref{fig:ETrho_2}), so the analysis of the
temperature evolution of the molecular charges for
the two $\beta^{\prime\prime}$ salts considered here will be based
on the $a~\nu_3$ mode in Raman spectra, which will complement that
based on the $b_1 \nu_{22}$, active in IR.\cite{girlando12}

\begin{figure}[ht]
\centering
\includegraphics[width=0.9\linewidth]{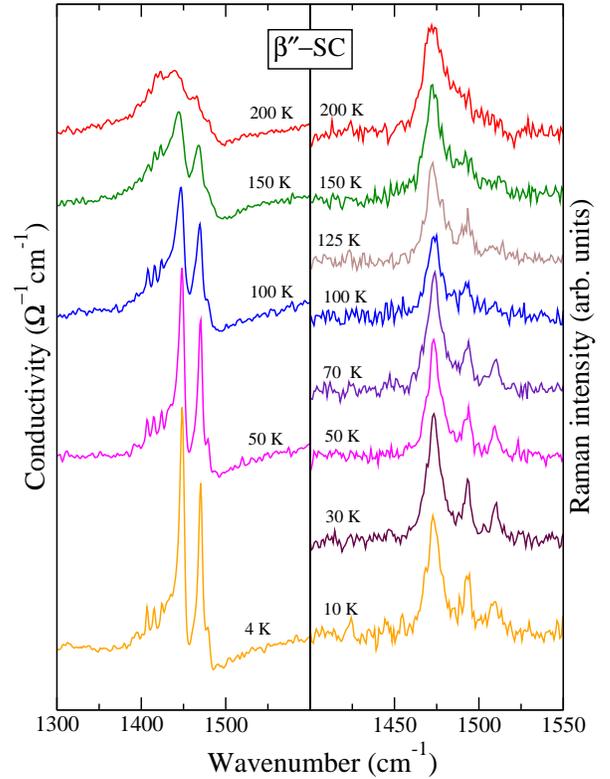}
\caption{Temperature dependence of infrared conductivity (left) and Raman
spectra (right) of $\beta^{\prime\prime}$-SC. The spectra are offset for clarity.
Notice the different wavenumber range in the IR and Raman spectra.}
\label{fig:betaSC:IRRam200}
\end{figure}

Fig.~\ref{fig:betaSC:IRRam200} compares the temperature evolution
of the infrared conductivity (left, from Ref.~\onlinecite{drichko09}) and
of the Raman spectra (right) of $\beta^{\prime\prime}$-SC in the spectral
region of interest. Whereas the ambient-temperature Raman spectrum (Fig. \ref{fig:ramanSC-300_10}) do not allow to clearly separate the contribution of the
$\nu_3$ and $\nu_4$ modes, at 200 K a broad feature
between 1485 and 1510 \cm\ can be unambiguously identified.
At room temperatures the $c$ polarized IR spectra present a
very broad band, centered around 1440 \cm, which is associated with
the $b_1 \nu_{22}$ mode and is indicative of charge
fluctuations.\cite{girlando12} At 200 K the 1440 \cm~IR band starts to show
a more complex structure (Fig.~\ref{fig:betaSC:IRRam200}, top left),
whereas at 150 K a doublet clearly
emerges both in IR and in Raman, signaling the beginning of some
sort of charge localization. By lowering the temperature,
the frequencies of the doublet remains basically unaltered, the separation of approximately 20~\cm\ corresponding to a charge difference $\Delta \rho$ of about 0.2
(Ref. \onlinecite{girlando12} and Fig.\ref{fig:ETrho_2}).
The intensities of the doublet increase with temperature,
a fact that can be better appreciated from the IR data.
The IR spectra also show that the broad band centered around 1440~\cm\ at
room temperature is still present at 4~K, with an intensity which
progressively decrease with lowering $T$. In Raman measurements, this is less evident,
due to the intrinsic weakness of the signal.

\begin{figure}[ht]
\centering
\includegraphics[width=0.9\linewidth]{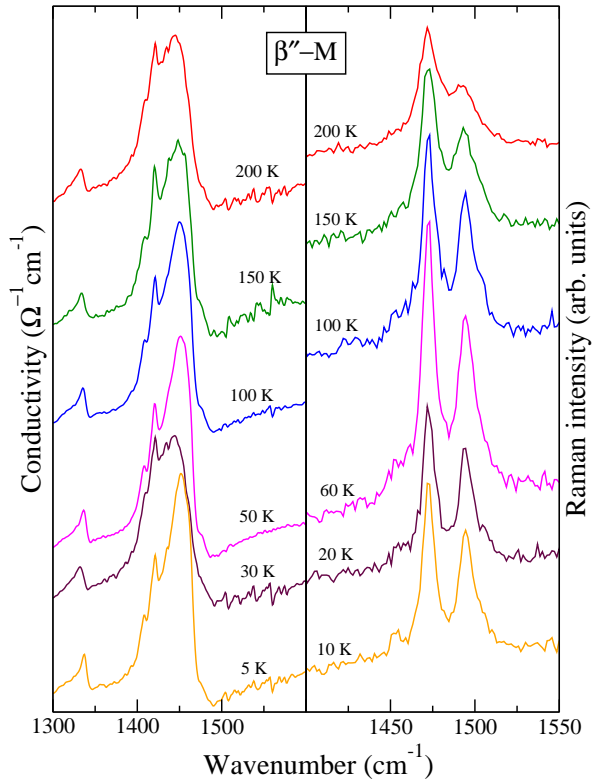}
\caption{Temperature dependence of infrared conductivity (left) and Raman spectra (right)
of $\beta^{\prime\prime}$-M. The spectra are offset for clarity. Note the
different wavenumber range in the IR and Raman spectra.}
\label{fig:betaM:IRRam200}
\end{figure}

The temperature-evolution of $\beta^{\prime\prime}$-M conductivity
and Raman spectra, shown in Fig.~\ref{fig:betaM:IRRam200},
differs markedly from that of $\beta^{\prime\prime}$-SC. In fact,
in IR conductivity the band associated to the $b_1 \nu_{22}$ mode,
around 1449 \cm, remains broad down to 5 K.
The situation is similar in the Raman spectra, where we can identify
the 1472~\cm~band with the $a~\nu_4$ mode, but where
at all temperatures we observe a single band, at about 1494~\cm,
to be associated with the $a~\nu_3$. Since the
x-ray structure indicate two inequivalent stacks
bearing different charges,\cite{ward00} it is clear
that in this case we have a persistent charge fluctuation regime,
that we shall now analyze in more detail.

\section{Analysis}

In Ref.~\onlinecite{girlando12} we have suggested that the frequency and band-shape of the
$b_1 \nu_{22}$ mode in the fluctuation regime can be understood by assuming the charge to ``jump''
stochastically between differently charged molecules.
Kubo has shown that in case of two chemical species
with some exchange between them and absorption energies relatively close,
the line shape depends on the difference in frequency between the two absorption energies
and the exchange rate. In case the exchange is slow, two distinct narrow bands are present.
As the exchange becomes faster the lines move closer together and widen.
In the limit of very fast exchange only one wide band is observed right
between the two levels.\cite{kubo69}

Kubo's ``two-states-jump model'' has been adapted to account for frequency and band-shape of the mode
in different regimes. The model can be actually used
both for IR and Raman, by substituting the Raman
intensity in place of the IR
oscillator strength.\cite{girlando12,yakushi12,Sedlmeier12}

The band-shape function is given by the real part of:
\begin{equation}
\mathcal{L}(\omega) = \frac{\mathcal{F}[(\gamma+2 v_{\mathrm{ex}}) - i (\omega-\omega_\mathrm{w})]}
{ \mathcal{R}^2-(\omega - \omega_A)(\omega -\omega_B)
- 2 i \Gamma(\omega-\omega_{\mathrm{av}})} \quad ;
\label{eq:bandshape}
\end{equation}
here $\mathcal{F} = f_A + f_B$, where $f_A$ and $f_B$ are
the oscillator strengths (Raman intensities) of the
bands of frequency $\omega_A$ and $\omega_B$ and common halfwidth $\gamma$.
The charge fluctuation velocity is $v_{\mathrm{ex}}$,
and $\Gamma = \gamma + v_{\mathrm{ex}}$ is the sum of the intrinsic width $\gamma$
and the exchange rate $v_{\mathrm{ex}}$, 	
$\mathcal{R}^2 = 2\gamma v_{\mathrm{ex}} + \gamma^2$.
Finally, the average and weighted frequency, $\omega_{\mathrm{av}}$ and
$\omega_{\mathrm{w}}$, are defined by:
\begin{equation}
\omega_{\mathrm{av}}=\frac{\omega_A+\omega_B}{2};~~~~~
\omega_{\mathrm{w}} = \frac{f_B \omega_A + f_A \omega_B}{f_A + f_B} \quad .
\label{eq:omega_av}
\end{equation}
If the transition rate $v_{\mathrm{ex}} \ll |\omega_A - \omega_B|/2$, Eq.~(\ref{eq:bandshape}) yields two separated bands located at $\omega_A$ and $\omega_B$, while if
$v_{\mathrm{ex}} \gg |\omega_A - \omega_B|/2$, the motional narrowing will
give one single band centered at the intermediate frequency $\omega_{\mathrm{av}}$.
On the other hand, when $v_{\mathrm{ex}} \approx |\omega_A - \omega_B|/2$
we shall observe one broad band whose spectral weight is
shifted towards the mode with higher oscillator strength.

Fitting the experimental spectra
with Eq.~(\ref{eq:bandshape}) will then
provide the frequencies $\omega_A$ and $\omega_B$, i.e.\
the corresponding ionicity $\rho$, the
ratio of the oscillator strengths, and
the velocity of the charge fluctuations.
In the IR range the fitting is constrained, as the frequencies and oscillator strengths
are bound to follow the calculated dependence on $\rho$.\cite{girlando11}

\begin{figure}[htp]
\includegraphics[width=\linewidth,clip=true]{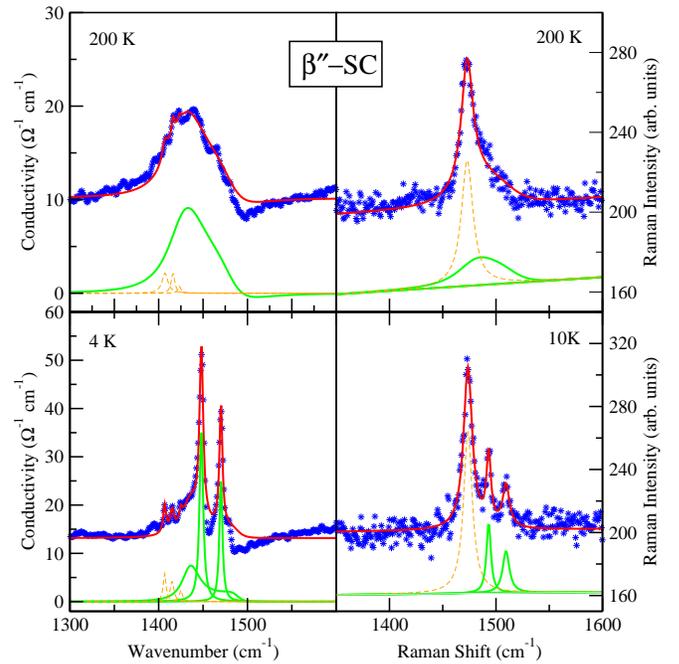}
\caption{Fit of the IR conductivity  (left) and of the Raman (right)
spectra of $\beta^{\prime\prime}$-SC at $T=200$~K and at the lowest attained temperature.
Blue crosses: experimental data; red line: overall spectral fit.
The dashed orange lines correspond to  Lorentzian fit of the bands
\textit{not} associated with the $b_1 \nu_{22}$ (IR) and $a~\nu_3$ (Raman). The green lines
are fits of the $b_1 \nu_{22}$ (IR) and $a~\nu_3$ (Raman) with the bandshape
given by Eq.~(\ref{eq:bandshape})}.
\label{fig:fitSC}
\end{figure}
The fit of the 200  and 5~K IR spectrum of $\beta^{\prime\prime}$-SC
is shown in the left panels of Fig.~\ref{fig:fitSC}.  At 200~K,
the single broad and somewhat structured band can be
satisfactorily fitted with three weak  Lorentzian bands associated
with charge-unrelated CH$_2$ bending modes (dotted orange lines
in the Figure),  and by Eq.~(\ref{eq:bandshape}) fluctuation model
relevant to the $b_1 \nu_{22}$ (green line). The fitting parameters
of the latter are reported in Tab.~\ref{table}.
The 4~K IR data are reproduced as the
superposition of two Lorentz oscillators, due to $b_1 \nu_{22}$ mode relevant to a static
CO state, and one Eq.~(\ref{eq:bandshape}) bandshape, related
to the surviving fluctuation regime, all indicated by the green line
in the bottom left panel of the figure. The static CO states correspond
to the Lorentzians bands  at  1448~\cm\ with $\rho_A = 0.58$ and 1470~\cm\ with $\rho_B = 0.42$, whereas the parameters of the fluctuating regime are reported in Table~\ref{table}.
\begin{table}[hb]
\centering
\caption{Two-state-jump model parameters obtained by the fit of
the IR and Raman bands due to the $b_1 \nu_{22}$ and $a_1 \nu_3$ modes,
respectively,  for $\beta^{\prime\prime}-$SC and $\beta^{\prime\prime}-$M salts.}
\begin{tabular}{l|rrcrrcrrcrr}
\hline \hline
\multicolumn{1}{c}{} & \multicolumn{5}{c}{$\beta^{\prime\prime}-$SC} & &
\multicolumn{5}{c}{$\beta^{\prime\prime}-$M} \\
\multicolumn{1}{c}{} & \multicolumn{3}{c}{IR} &\multicolumn{2}{c}{Raman} & &
\multicolumn{3}{c}{IR} &\multicolumn{2}{c}{Raman} \\
\vspace*{1mm}
$T$~(K) & 200  & 4  &~& 200  &10  & & 200  & 5 &~ & 200  & 10  \\
\cline{2-6} \cline{8-12}
$\Delta\omega$~(\cm)    & 72    & 56    &~ &   57   & -- &~~ & 55  &  45    &~&   40  & 25 \\
$\Delta\rho$                 & 0.52 & 0.40 &~&   0.52 & -- &~~ & 0.40 & 0.32 &~& 0.38 & 0.30 \\
$v_{\mathrm{ex }}$~(\cm)& 22   &   11   &~&    21   & -- &~~&   24  & 22    &~&   26  & 25 \\
\hline \hline
\end{tabular}
\label{table}
\end{table}

In the 200~K Raman spectrum (right upper panel of Fig.~\ref{fig:fitSC})
the $a~\nu_4$ mode peaks at 1472~\cm; we have fitted the band by a simple Lorentzian
since even in the fluctuation regime the exchange velocity, as estimated from the IR data,
is much less than the frequency difference between the two differently charged
states (Fig.~\ref{fig:ETrho_2}). Once this contribution is subtracted, the high frequency side of the
band, due to the $a~\nu_3$, can be fitted by Eq.~(\ref{eq:bandshape}). The
obtained parameters are consistent with those extracted from the IR spectrum and
are summarized in Tab.~\ref{table}.
In the $T=10$~K Raman spectrum
(lower right panel of Fig.~\ref{fig:fitSC}) we can identify three bands.
The low-frequency peak corresponds to the $a~\nu_4$ mode of the charge-rich molecules.
The two other Raman bands are assigned to the $a~\nu_3$ mode of the charge-rich
and charge-poor  BEDT-TTF in a CO frozen state. They can be analyzed in terms of simple Lorentzian oscillators
centered around 1493 and 1509~\cm, corresponding to $\rho = 0.43$ and 0.57,
respectively. The findings are in perfect agreement with the results obtained
from the analysis of the IR $b_1 \nu_{22}$ mode. Given the weak intensity
of the spectrum, it is not possible in this case to identify bands which could
be attributed to the surviving fluctuation regime.

\begin{figure}[htp]
\centering
\includegraphics[width=\linewidth]{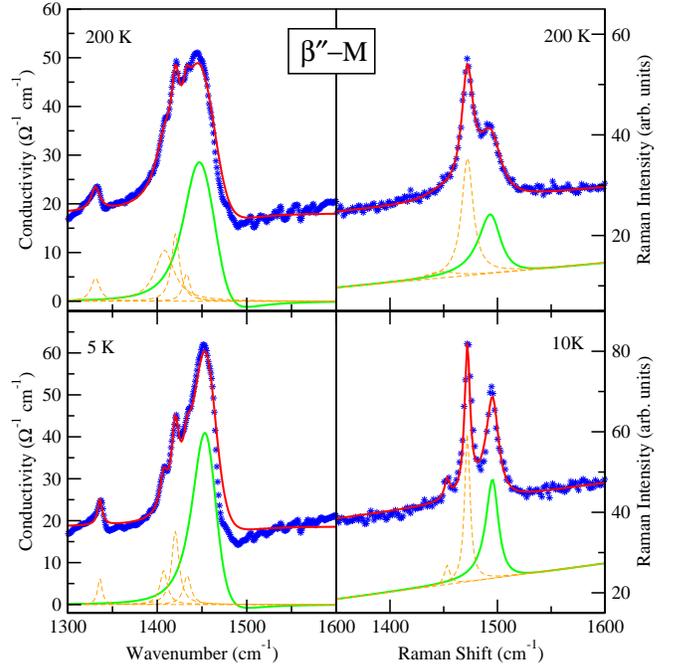}
\caption{Fit of the IR conductivity  (left) and of the Raman (right)
spectra of $\beta^{\prime\prime}$-SC at 200 K and at the lowest attained $T$.
Blue crosses: Experimental data; Red line: Overall spectral fit.
Dashed orange lines:  Lorentzian fit of bands \textit{not} associated
with the $b_1 \nu_{22}$ (IR) and $a~\nu_3$ (Raman). Green lines:
Fit of the $b_1 \nu_{22}$ (IR) and $a~\nu_3$ (Raman) with the bandshape
given by Eq.~\ref{eq:bandshape}}.
\label{fig:fitM}
\end{figure}

For the metallic compound  $\beta^{\prime\prime}$-M
the fits of the 200 and 10~K IR and Raman spectra are presented
in Fig.~\ref{fig:fitM} in the spectral region of the charge sensitive
modes. The 200~K spectra look very similar to the corresponding ones
of $\beta^{\prime\prime}$-SC, and can be described with the two-state jump model.
Here we had to subtract the charge-unrelated bands, either connected to
the CH$_2$-bending mode
in the IR spectrum or the $a~\nu_4$ feature observed in the Raman spectrum (dashed orange lines).
The fit parameters for the IR and Raman data are reported in Tab.~\ref{table}.
When commenting Fig.~\ref{fig:betaM:IRRam200}
we have already noted that in $\beta^{\prime\prime}$-M the fluctuation regime
is retained down to lowest temperatures. The analysis of the IR and Raman spectra, performed
in the same fashion as for the 200~K spectra, provides a consistent picture:
the obtained parameters are reported in Tab.~\ref{table}.
The comparison between the fit parameters at the two different temperature reveals
that the variation in the spectra upon lowering $T$ is more due to a change
in the ground state ionicity than to a change in $v_{\mathrm{ex}}$.

\section{Discussion}
\begin{figure}
\centering
\includegraphics[width=0.8\linewidth]{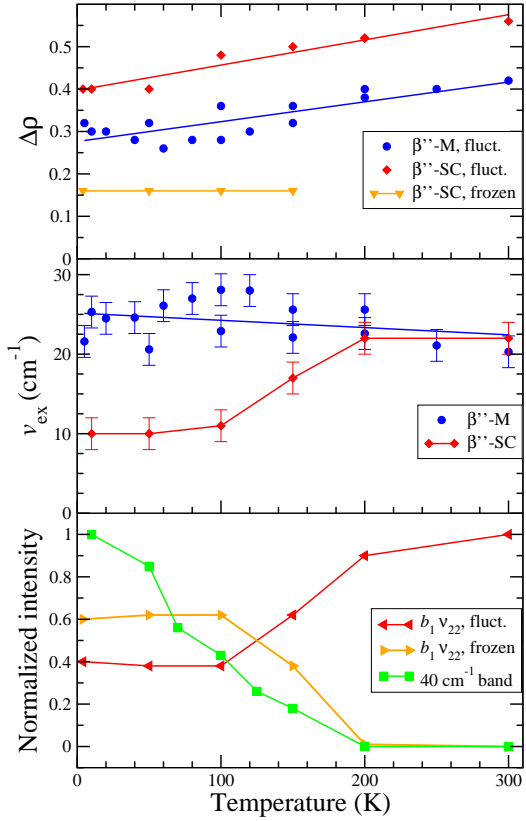}
\caption{Top and middle panels: Charge difference $\Delta \rho$ and
exchange velocity $v_{\rm ex}$ of $\beta^{\prime\prime}$-SC and
$\beta^{\prime\prime}$-M  as a function of temperature.
The error bars in the middle panel are an estimate of the fitting uncertainty,
based on the scattering of the various points. Bottom panel: normalized
intensity of the $b_1 \nu_{22}$ (fluctuating and frozen CO components)
band, and of the band associated with a lattice phonon around 40~\cm.}
\label{fig:deltarho_Vex}
\end{figure}

The simultaneous analysis of the IR and Raman spectra of $\beta^{\prime\prime}$-SC
and $\beta^{\prime\prime}$-M gives a quite consistent picture.
At room temperature both salts are in a charge-order fluctuation
regime. The A and B stacks bear a different amount of charge, and
fluctuations occur between these ``stripes'' aligned along the $a$-axis.
Upon lowering the temperature to 5-10 K,
the spectra of $\beta^{\prime\prime}$-M and $\beta^{\prime\prime}$-SC
evolve distinctively.
This difference in the temperature-dependence is illustrated in detail
in the top and middle panel of Fig.~\ref{fig:deltarho_Vex},
where we report the parameters which best characterize the charge order,
the disproportionation $\Delta \rho$ and  the exchange velocity $v_\mathrm{ex}$.
For the metallic salt $\beta^{\prime\prime}$-M, the fluctuation regime persists down
to the lowest temperature; the exchange velocity remains essentially unchanged
(within the fitting uncertainties) and the charge distribution becomes slightly
more uniform. In the SC salt  $\beta^{\prime\prime}$-SC,
a ``frozen'' CO state starts to emerge below 200~K, well above the superconducting phase
at $T_c=5$~K;
this state coexists with the fluctuation regime down to the lowest temperatures.

The temperature evolution of the spectra (Fig.~\ref{fig:betaSC:IRRam200})
suggests that the  relative weight of the frozen CO regime, whose $\Delta \rho$
is constant with $T$, increases at the expenses of the fluctuating regime.
This is quantitatively shown in the bottom panel of Fig.~\ref{fig:deltarho_Vex},
where the temperature dependence of the normalized intensities of the corresponding bands
is plotted.
The normalization is performed with respect to the total intensity of the bands,
which is approximately constant with temperature.
The bands corresponding to the frozen CO regime start to borrow intensity
from those corresponding to the fluctuating regime mainly in the
range between 200 and 100~K.
Below this temperature, the relative intensities remains approximately constant, and
at the superconducting temperature the fluctuating
regime is still present.
Indeed,  $v_\mathrm{ex}$ of $\beta^{\prime\prime}$-SC (middle panel
of Fig.~\ref{fig:deltarho_Vex}) halves between 200 and 100 K, but does
not go to zero, which would imply a complete metal to insulator transition.

It is instructive to connect the present data with complementary
measurements on the two salts. The x-ray analysis indicates that
a charge disproportionation is already present at 300~K,\cite{geiser96,ward00}
but the existence of the charge fluctuation regime
makes both salts two-dimensional metals; actually ``bad metals'' as for most
BEDT-TTF salts. By lowering $T$,
the resistivity of $\beta^{\prime\prime}$-SC exhibits a change in
slope around~150 K, where the spectral signatures of a frozen CO clearly appear,
and the fluctuations starts to slow down.
In contrast, no anomalies are observed in the temperature-dependent
resistivity of $\beta^{\prime\prime}$-M.\cite{glied09}
\begin{figure}
\centering
\includegraphics[width=0.9\linewidth]{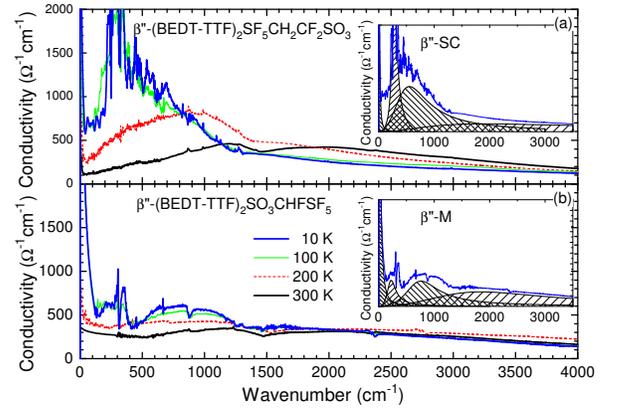}
\caption{Temperature dependence of the $b$-polarized in-plane conductivity spectra of $\beta^{\prime\prime}$-SC
and $\beta^{\prime\prime}$-M measured in a wide frequency range. The inset visualizes the different
contributions to the low-temperature conductivity (from Ref.~\onlinecite{kaiser10}).}
\label{fig:betacond_inplane}
\end{figure}

As shown in Fig.~\ref{fig:betacond_inplane},  the optical conductivity within
the $ab$ plane contains three main
features which are common to both salts:\cite{kaiser10}
\begin{itemize}
\item A broad band around 2000-3000 \cm, associated with site-to-site transitions within fluctuating CO patterns;
\item a narrow Drude peak describing the coherent-carrier response that is also responsible for the dc transport and superconductivity;
\item and a so-called ``charge fluctuation band'' observed for quarter-filled organic conductors
typically around $500-700$~\cm.\cite{merino06,drichko06}
\end{itemize}
In addition there are two features observed in the in-plane conductivity spectra, which distinguish $\beta^{\prime\prime}$-SC from $\beta^{\prime\prime}$-M, and which start to develop below
approximately 150~K when the charge localization begins:\cite{kaiser10}
\begin{itemize}
\item
A phonon mode at $30-40$~\cm\ that is only observed for the electric vector polarized
perpendicular to the $b$-axis (therefore roughly parallel to $a$, cf.
Fig.~\ref{fig:beta_SC1}). The temperature evolution of its
intensity (normalized to the intensity at the lowest temperature)
is reported in the lower panel of Fig.~\ref{fig:deltarho_Vex} (squares, green line).
\item
An additional broad feature
appears around 300~\cm\ in the optical spectrum polarized along the $b$ axis.
The temperature evolution of this band is difficult to follow,
since the band changes shape and position (in addition to the intensity).
A similar mode may also be present in the  $\beta^{\prime\prime}$-M salt;
however its intensity is much weaker and it remains a more or less
temperature independent.
\end{itemize}

The interpretation of this distinctive second feature remains uncertain;\cite{kaiser10}
it is not clear whether and how it is associated with the presence of the mixed phase,
containing fluctuating and frozen CO.
On the other hand, the bottom panel of Fig.~\ref{fig:deltarho_Vex}
demonstrates that the intensities of the $30-40$~\cm\ band and the $b_1 \nu_{22}$ mode that is associated  to the frozen CO regime below $T=200$~K increase simultaneously.
This is a strong indication that the two features have the same origin.
Hence we associate the low-frequency mode with oscillations of the frozen CO.
Since this mode is polarized along the $a$-axis, namely along the stacks,
the frozen CO is established along the stack.
In other words the inversion symmetry present in the stacks is actually broken,
and the charge disproportionation involves the A and A$^{\prime}$ (and in analogy B and B$^{\prime}$)
molecules along the stack (Fig.~\ref{fig:beta_SC1}).
This symmetry breaking is not detected by the x-ray analysis below
$T=150$~K,\cite{geiser96} but it is well known that x-ray detects long range order only,
whereas vibrational spectroscopy probes the local environment. In this
interpretation, the low-frequency phonon would be associated with
oscillations of a charge-ordered string along the $a$-axis, whereas we recall
that the fluctuations presumably occur between the charge-ordered stripes on
different stacks, namely, along the $b$-axis (Fig.~\ref{fig:beta_SC1}).

In the light of the above data, we suggest that below approximately 200~K
two different, almost orthogonal, types of charge order start to develop
in $\beta^{\prime\prime}-$SC:
a fluctuating CO between A and B stacks, leading to the broad bands interpreted
in terms of Eq. (\ref{eq:bandshape}), and frozen
CO strings along the stacks, leading to the low-frequency mode at
$30-40$~\cm. This picture is strongly reminiscent of the
coexistence of different sorts of charge modulation observed by diffuse x-ray  in $\theta$-(BEDT-TTF)$_2$Cs$M^\prime$(SCN)$_4$ ($M^\prime$ = Co, Zn).\cite{nogami99}
Udagawa and Motome\cite{udagawa07} have attributed this coexistence to the competition between nearest-neighbor electron-electron interaction, causing a regular charge order,
and Fermi surface nesting, leading to a commensurate charge-density wave.
A similar scenario might be applicable to the present case,
since the frozen CO appears to be related to an intermolecular phonon.
Diffuse x-ray investigation as a function of temperature on both $\beta^{\prime\prime}$-SC
and $\beta^{\prime\prime}$-M amended by appropriate modeling is required to confirm this idea.

It is instructive to take into consideration two other salts in the series
$\beta^{\prime\prime}$-(BEDT-TTF)$_2$SF$_5$RSO$_3$, i.e.
$\beta^{\prime\prime}$-(BEDT-TTF)$_2$SF$_5$CH$_2$SO$_3$ (hereafter $\beta^{\prime\prime}$-I),\cite{ward00}
and $\beta^{\prime\prime}$-(BEDT-TTF)$_2$SF$_5$CHFCF$_2$SO$_3$ (hereafter $\beta^{\prime\prime}$-MI). \cite{olejniczak99,jones00,schlueter01,garlach02}  The two salts are isomorphous
with $\beta^{\prime\prime}$-SC and $\beta^{\prime\prime}$-M, but have very distinct
electrical properties: $\beta^{\prime\prime}$-I is semiconductor already at room temperature,
whereas $\beta^{\prime\prime}$-MI has a metal-to-insulator transition around $T=180$~K.
Unfortunately, infrared data have been collected only in the conducting plane,
and Raman spectroscopy has been measured at room temperature only.
From the optical conductivity\cite{olejniczak09}
and from a reanalysis of the Raman data presented in Ref.~\onlinecite{ward00},
we deduce that $\beta^{\prime\prime}$-I is in a full CO state, with $\Delta \rho
\approx 0.5$ to 0.6. On the other hand, the room temperature
Raman spectrum of $\beta^{\prime\prime}$-MI in the charge sensitive region
is practically identical to that of  $\beta^{\prime\prime}$-M,\cite{schlueter01}
suggesting a fluctuating CO regime with very similar $\Delta \rho$.
The metal-insulator transition of $\beta^{\prime\prime}$-MI has been ascribed
to electron localization due to anion ordering. It is plausible
to assume that the anion ordering also induces a CO on the cation.

To summarize, at room temperature the four salts of the
$\beta^{\prime\prime}$-(BEDT-TTF)$_2$SF$_5R$\,SO$_3$ family all display a
charge disproportionation $\Delta \rho$ between 0.4 and 0.6,
most probably between the A and B stacks (cf.\ Fig.~\ref{fig:beta_SC1}).
$\beta^{\prime\prime}$-I is in a fully charge-ordered state, whereas
the other three salts are in a charge-fluctuation regime, with metallic
conductivity. By lowering the temperature, $\beta^{\prime\prime}$-M
remains in the fluctuating regime, whereas $\beta^{\prime\prime}$-MI undergoes
a phase transition around $T=180$~K accompanied by charge localization.
Finally, below 200~K $\beta^{\prime\prime}$-SC exhibits a ``mixed
phase'', with the coexistence of fluctuating and frozen
regimes. This coexistence persists down to the superconducting phase.

\begin{figure}[floatfix]
\centering
\includegraphics[width=0.85\linewidth]{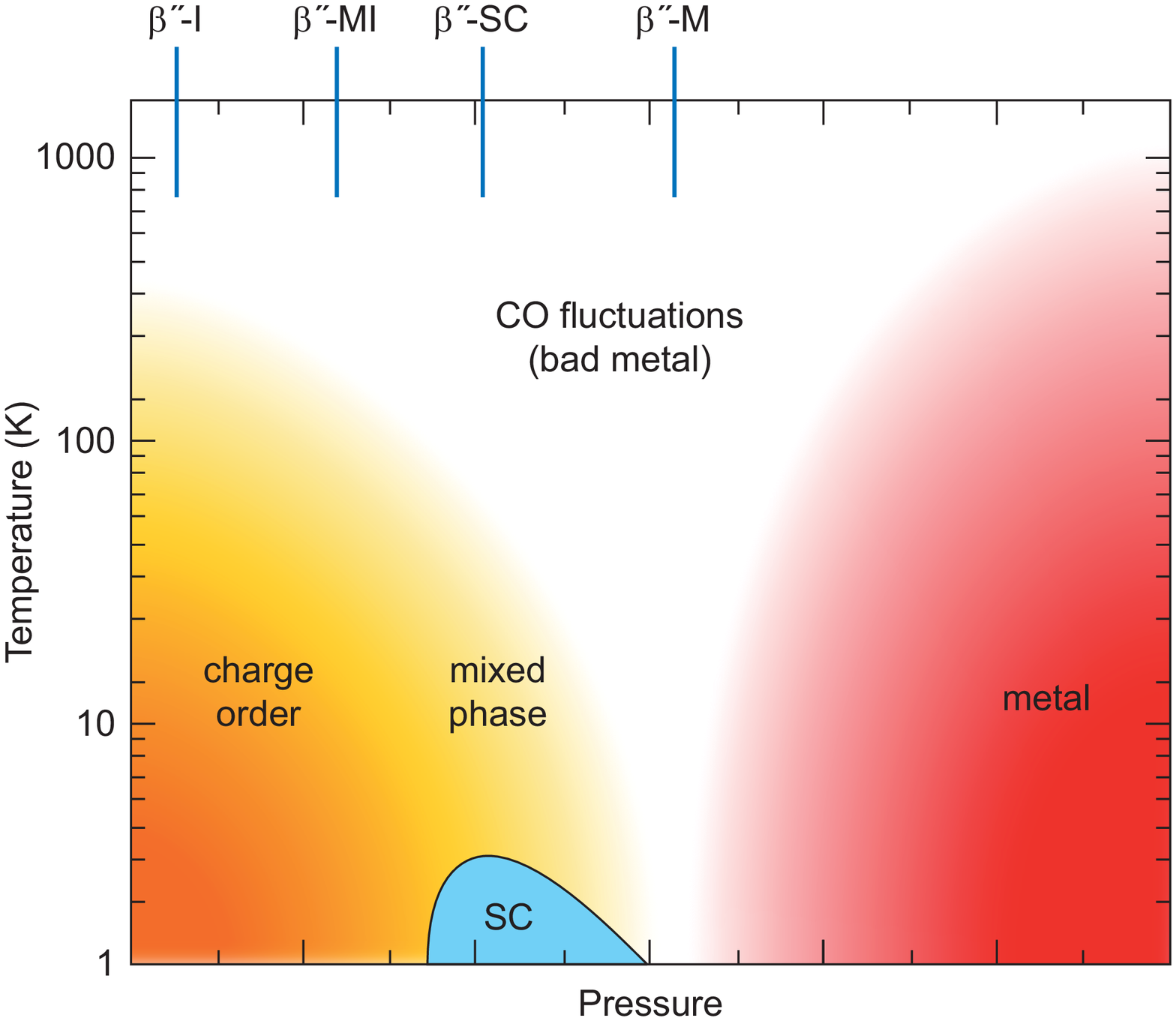}
\caption{Tentative phase diagram for the members of the $\beta^{\prime\prime}-$(BEDT-TTF)$_2$SF$_5R$\,SO$_3$ family.
$\beta^{\prime\prime}$-(BEDT-TTF)$_2$SF$_5$CH$_2$SO$_3$ (denoted $\beta^{\prime\prime}$-I)
is a charge-ordered insulator,  $\beta^{\prime\prime}$-(BEDT-TTF)$_2$SF$_5$CHFCF$_2$SO$_3$
($\beta^{\prime\prime}$-MI) undergoes a metal-insulator transition at 180~K,
$\beta^{\prime\prime}$-(BEDT-TTF)$_2$SF$_5$CH$_2$CF$_2$SO$_3$  ($\beta^{\prime\prime}$-SC)
is a superconductor at $T_c=5$~K,
and $\beta^{\prime\prime}$-(BEDT-TTF)$_2$SF$_5$CHFSO$_3$ ($\beta^{\prime\prime}$-M) remains metallic
down to low temperatures.}
\label{fig:phasediagra1}
\end{figure}
Based on the above scenario, we propose the empirical phase
diagram presented in Fig.~\ref{fig:phasediagra1}. The pressure
axis refers to chemical pressure, and could be replaced by the intermolecular
Coulomb interaction $V$, running in opposite direction.
This diagram can be compared with that theoretically proposed
for the $\beta^{\prime\prime}$ and the $\theta$ (BEDT-TTF)$_2X$
structures\cite{merino01} and also applicable to the $\alpha$-(BEDT-TTF)$_2X$  compounds.\cite{Dressel03,Drichko06}
Our empirical diagram is very similar, the conductivity of the metallic phase is
typical of a ``bad metal'', and presumably due to charge fluctuations
between the BEDT-TTF units.  However, we also
find that between the metallic phase and the charge-ordered insulating phase,
there is a region of coexistence, where the system is still metallic,
but regions of frozen charge order develop. The region of coexistence is rather
intriguing, and of course cannot be predicted within mean field theories.
On the other hand, the coexistence is not due to macroscopic
domains at the borderline between two phases, since the temperature
interval is rather large, and perfectly reproduced in different runs
and measurements. As we have already stated, the coexistence is likely  due
to the competition between two different interactions, driving towards
different charge-ordered states.

\section{Conclusions}
In a comprehensive study we have compared the temperature dependence of the Raman and
infrared active charge-sensitive vibrations of BEDT-TTF in $\beta^{\prime\prime}$-SC and
$\beta^{\prime\prime}$-M salts. We could delineate the interplay of charge order, charge fluctuations and superconductivity in these two-dimensional quarter-filled electron systems.
Two distinct charge-ordered states could be identified in $\beta^{\prime\prime}$-SC:
one fluctuating and the other static; they are almost orthogonally polarized and
coexist below 200~K down to the critical temperature $T_c=5$~K.
This ``mixed phase'', not found in $\beta^{\prime\prime}$-M, is likely the results
of the competition/cooperation of different interactions.
It then becomes clear that very weak additional interactions
may tune the system towards one or the other ground state. In this scenario, it is uncertain whether or not the ``mixed phase'' is essential to the development of superconductivity, since so far this phase has been found only in $\beta^{\prime\prime}$-SC, and not in other superconductors
with either $\beta^{\prime\prime}$ or in the similar $\theta$ or $\alpha$ structure.
But of course it has not been expressly looked for.

\section*{Acknowledgments}
The work in Parma has been supported by the Italian Ministry
of University and Research (M.I.U.R.) under the
project  PRIN-2010ERFKXL. We would like to thank
the Deutsche Forschungsgemeinschaft (DFG) for financial
support. N. Drichko acknowledges a support of
Margarete von Wrangell Habilitationstipendium.
J. A. Schlueter acknowledges support from the Independent Research/Development program while serving at the National Science Foundation.

\end{document}